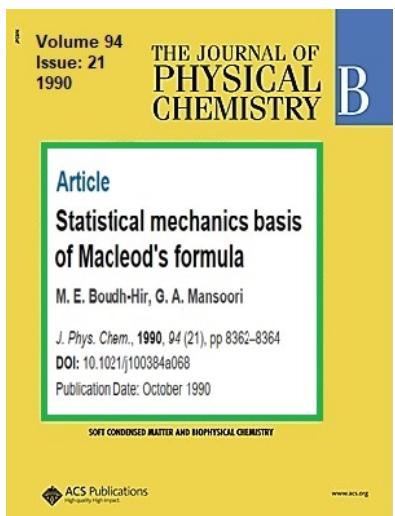

# Statistical Mechanics Basis of Macleod's Formula

**M.-E. BOUDH-HIR** and **G.A. MANSOORI**[*]
**University of Illinois at Chicago**
(M/C 063) Chicago, IL 60607-7052, USA

---

(*) Corresponding Author, email: mansoori@uic.edu

**Abstract**

In this work the theoretical basis for the famous formula of Macleod, relating the surface tension of a liquid in equilibrium with its own vapor to the one-particle densities in the two phases of the system, is derived. Using the statistical-mechanical definition of the surface tension, it is proved that this property is, at the first approximation, given by the Macleod formula.

**1. Introduction**

Macleod's empirical formula[1] is probably one of the simplest formulae known in thermodynamics. It expresses the surface tension of a liquid in equilibrium with its own vapor as a function of the one-particle densities, $\rho_l$ and $\rho_v$, in the liquid and vapor phases as follows:

$$\gamma = K (\rho_l - \rho_v)^4, \tag{1}$$

where $K$ is a constant independent of the system temperature but dependent on the nature of the fluid under consideration. Other forms for this equation have been proposed. Here we just mention the version given by Sugden[2] who wrote this equation in the form:

$$P = MK^{1/4} = M\gamma^{1/4}/(\rho_l - \rho_v). \tag{2}$$

In this version of Macleod's formula, $M$ is the molar mass and $P = MK^{1/4}$ is what Sugden called the parachor. For more details, one may see reference 3. It has been shown that $P$ is nearly constant for variety of fluids and within wide ranges of temperature. It follows that this equation relates the volumes of the two phases of the system to its surface tension. It may therefore be considered, from some point of view, as an equation of state for the interface. This version has been presented as a mean to compare, at constant surface tension, the molar volumes for two different substances[2].

Equation (1) has two important qualities, among others, which have made it so popular: first, the extreme simplicity of its analytical form makes the calculations easy to perform. Second, this formula works very well for many substances and over a wide range of temperature. This quality is, of course, very important. Therefore, there are strong reasons for us to be interested in finding the statistical-mechanical basis of this formula. This knowledge is indeed important: *(i)* from the fundamental point of view, it is interesting to understand how such an equation may be derived, *(ii)* from the practical point of view, this knowledge may help us to modify this equation in order to improve its accuracy and/or to extend its use to the calculation of the interfacial tension.

A number of efforts have been made in the past to justify the success of this formula from the theoretical basis[4-7]. Two different methods have been used: the first starts with the classical thermodynamic equation relating the surface tension, $\gamma$, to the surface internal energy, $u$,

$$u = S\,(\gamma - T\,\partial_T \gamma), \qquad (3)$$

in which, $S$ and $\partial_T$ denote the surface area and the differentiation with respect to the temperature, respectively. The second, whose details are given later, begins with the statistical-mechanical definition of the surface tension. Using the first method[4,5], it has been shown that the surface tension is proportional to the power four of the difference in densities, $(\rho_l - \rho_v)$. However, the use of certain simplifying statistical-mechanical approximations[6,7] leads to results different from equation (1).

Here, starting with the statistical-mechanical definition of the surface tension, we prove that this property, as the first approximation, is given by the Macleod formula. As a result, the law in power four of the difference in densities is obtained. However it is shown that the constant, $K$, which depends on the nature of the given fluid is not rigorously independent of temperature of the system under consideration. The only simplification made is that the particles are considered interacting via a generalized additive pairwise potential. This interaction may however depend on the respective positions and orientations of the molecules (i.e., the particles are not assumed to be spherical and the potential is not necessarily a radial function).

## 2. Model and theoretical developments

Let us consider a system of identical particles interacting via a pairwise potential, $w(i,j)$. In the general case, the interaction between two particles $i$ and $j$ maybe expressed as a function of the respective orientations, $\Omega_i$ and $\Omega_j$, of these particles and the vector, $r_{ij}$, joining their centers. We have:

$$w(i,j) = w(r_{ij}, \Omega_i, \Omega_j). \qquad (4)$$

The surface tension is defined as (see for instance chapter 4 of reference 6):

$$\gamma = (\partial_S F)_{V,T,\mu} = (1/2) \iint \rho(1,2)\,(\partial_S\,w(1,2))_V\,d1\,d2. \qquad (5)$$

Here, $i = 1, 2$ denotes the generalized coordinates of the $i$-th particle (i.e., $i \equiv (r_i, \Omega_i)$), $\mu$ is the chemical potential, $\partial_S$ stands for the differentiation with respect to the surface area, $F$ is the Gibbs free energy of the system, $V$ and $T$ are volume and temperature of the system, respectively, and $\rho(1,2)$ represents the two-particle density defined by:

$$\rho(1,2) = \Xi^{-1} \sum_N (N(N-1) z^N/N!) \, {}^o\!\!\int_{\leftarrow N-2 \rightarrow} \exp\{-\beta \sum_{i,j} w(i,j)\} \, d\{N-2\}. \quad (6)$$

$\Xi$ is the grand partition function of the system and $d\{N-2\}$ means that the integration is to be performed over the positions and the orientations of the $N-2$ particles.

To perform the differentiation with respect to the surface area, one may use the following technique (see for example, reference8):

$$\mathbf{r} = (x, y, z) = (S^{1/2} x', S^{1/2} y', V z'/S). \quad (7)$$

Then equation (5) takes the form:

$$\gamma = (1/4S) \iint \rho(1,2) (\mathbf{r}_{12} - 3\mathbf{z}_{12}) \cdot \partial_{\mathbf{r}_{12}} w(1,2) \, d1 \, d2 = (1/4S) \iint \rho(1,2) \mathbf{r}_{12}^* \cdot \partial_{\mathbf{r}_{12}} w(1,2) \, d1 \, d2 \quad (8)$$

in which, the vector function, $\partial_{\mathbf{r}_{12}} w(1,2)$, is defined as having the components $\partial_{x_{12}} w(1,2)$, $\partial_{y_{12}} w(1,2)$ and $\partial_{z_{12}} w(1,2)$. For homogeneous systems, the three directions $x$, $y$ and $z$ are equivalent and so, $\gamma$ is zero as it should be.

The most direct way to derive the system thermodynamic properties from its Gibbs free energy is the use of the diagrammatic expansion theory. This technique may be used here. One has:

* First, to decompose the activity, $z$, and the Mayer function, $f$, into two parts as follows:

$$z = (2\pi mkT/h^2)^{3/2} e^{\beta\mu} = z_c + z^*, \quad (9a)$$

$$f(i,j) = e^{-\beta w(i,j)} - 1 = f_c(i,j) + f^*(i,j). \quad (9b)$$

The subscript $c$ denotes the values of these functions at the critical temperature while the star stands for their deviations from the critical values.

* Second, to proceed to the usual topological reductions.

However, because: *(i)* the correlation functions on the interface become long-ranged functions and so, the calculations have to be done very carefully, and *(ii)* in the present work the full development is not needed, the use of other methods is more convenient.

It has been shown that the surface tension, $\gamma$, may be related to the direct correlation function[9-11]. We have:

$$\gamma = (kT/4S) \iint c(1,2) \, \partial_{z_1}\rho(1) \, \partial_{z_2}\rho(2) \, (r^2_{12} - z^2_{12}) \, d1 \, d2. \quad (10)$$

The equivalence of the two equations (8) and (10) has first been proven for simple fluids[12] then it was extended to the case of molecular fluids[13]. The discrepancy of the use of equation (10) is that it makes arise the direct correlation function about which we know almost nothing on the interface. This equation however presents the advantage that it introduces a function remaining short-ranged for two-phase systems (see chapters 7 and 9 of reference 6). The one-particle density satisfies[14,15]:

$$\rho(i) = z \exp(c(i)), \quad (11)$$

where $c(i)$ denotes the one-particle direct correlation function and $z$ is the activity of the particles given in equation (9a).

Differentiating equation (11) with respect to $z_i$, we obtain:

$$\partial_{z_i} \rho(i) = \rho(i) \int c(i,j) \, \partial_{z_j} \rho(j) \, dj, \qquad (12)$$

which leads after integration over $z_i$ to:

$$\rho(i) = [(\rho_l - \rho_v)/2\Omega] \exp\left\{ \int_{z_0}^{z_i} dz'_i \int\int c(i,j) \, \partial_{z_j}\rho(j) \, dr_j \, d\Omega_j \right\}$$

$$= [(\rho_l - \rho_v)/2\Omega] \exp\left\{ \int_{z_0}^{z_i} dz'_i \int\int C(z'_i, z_j, \Omega_i, \Omega_j) \, \partial_{z_j}\rho(j) \, dz_j \, d\Omega_j \right\}. \qquad (13)$$

In this equation, $z_0$ denotes the position of the plan located in the region where liquid and vapor coexist and such that the one-particle density averaged over the orientation of the particle is equal to $(\rho_l - \rho_v)/2$.

It is now easy to see that the first term of $\partial_{z_i}\rho(i)$ considered as a function of the quantity $(\rho_l - \rho_v)$ is proportional to $(\rho_l - \rho_v)^2$. Therefore, using equations (10) and (13) the surface tension can be written in the form:

$$\gamma = K(\rho_l - \rho_v)^4, \qquad (14)$$

when only the first term non-zero is retained. The constant $K$ depends however on the temperature of the system. Far from the critical temperature, $K$ behaves as $T^{-4}$. It follows that the change in $K$ with respect to the temperature is small because of the fact that for a given liquid/vapor system the change in temperature is not very important compared to the initial temperature.

Near the critical temperature, the surface tension remains given by equation (14). Nevertheless, the constant *K* becomes more sensitive to the temperature. To treat this case, we will start with equation (10) in which the one-particle density is expanded around its critical value $\rho_c$. We have:

$$\rho(i) = (\rho_c z/z_c) \cdot exp(\Delta c(i)), \tag{15}$$

where $\Delta c(i)$ denotes $c(i) - c_c(i)$. Equations (10) and (15) together lead to:

$$\gamma = (kT/4S)(\rho_c z/z_c)^2 \int\int \partial_{z_1} e^{\Delta c(1)} \partial_{z_2} e^{\Delta c(2)} c(1,2)(r^2_{12} - z^2_{12}) d1 d2. \tag{16}$$

To solve this equation, the dependence of the correlations functions on the positions and orientations of the particles, the temperature of the system and their functional dependence on the one-particle density should be known. Being the problem to study, the lowest order (non-zero) is sufficient. In order to find this term, a method quite similar to the familiar charging process, employed in the Coulombic potential case, will be used. We therefore define the parameters, $\xi$ and $\theta$, as follows:

$$\beta(\xi) = \beta_c/(1-\xi\theta), \tag{17}$$

$$\theta = (T_c - T)/T_c. \tag{18}$$

$\Delta c(i; \xi)$ is simply defined by:

$$\Delta c(i; \xi) = c(i; \beta(\xi)) - c(i; \beta_c), \tag{19}$$

and it satisfies:

$$\Delta c(i) = \int_0^1 \partial_\xi \Delta c(i; \xi) \, d\xi. \tag{20}$$

It is clear that $\xi=0$ corresponds to the system at its critical temperature, $T_c$, while $\xi=1$ is associated with the system at the temperature of interest, $T$. $\Delta c(i; \xi)$ depends on $\xi$ through the Mayer function, $f(i,j; \xi)$, and the one-particle density, $\rho(i; \xi)$. It follows that:

$$\Delta c(i) = \int_0^1 \left\{ \iint \delta_{f(j,k;\xi)} c(i;\xi) \, \partial_\xi f(j,k;\xi) \, dj \, dk + \int \delta_{\rho(j;\xi)} c(i;\xi) \, \partial_\xi \rho(j;\xi) \, dj \right\} d\xi. \qquad (21)$$

Here, $\delta_x$ stands for the functional differentiation with respect to $x$. In the first differentiation, one of the arguments $j$ and $k$ may be identical to $i$, however this is not possible in the second. Taking this remark into account, equation (21) can be written in the following form:

$$\Delta c(i) = \int_0^1 d\xi \left\{ \int c(i,j;\xi) \rho(j;\xi) \partial_\xi f(i,j;\xi) \, dj + \tfrac{1}{2} \iint c(i,j,k;\xi) \rho(j;\xi) \rho(k;\xi) \partial_\xi f(j,k;\xi) \, dj \, dk \right.$$
$$\left. + \int c(i,j;\xi) \partial_\xi \rho(j;\xi) \, dj \right\}. \qquad (22)$$

When the term containing the three-particle direct correlation function, which evidently has a higher order contribution and so, it is not needed for this purpose, is neglected, this equation becomes:

$$\Delta c(i) = \rho_c \int_0^1 d\xi \int c(i,j;\xi) \left\{ [z(\xi)/z_c] \partial_\xi f(i,j;\xi) e^{\Delta c(j;\xi)} + \partial_\xi [(z(\xi)/z_c) e^{\Delta c(j;\xi)}] \right\} dj. \qquad (23)$$

Performing the differentiation with respect to $\xi$ and retaining only the lowest order non-zero in $\theta$, we may write:

$$\Delta c(i) = \rho_c \int_0^1 d\xi \int c(i,j;\xi) \left\{ \theta [-\beta_c v(i,j) e^{-\beta_c w(i,j)} + \beta_c \mu - 3/2] e^{\Delta c(j;\xi)} + \partial_\xi e^{\Delta c(j;\xi)} \right\} dj. \qquad (24)$$

Using the iteration process, it can be shown that the contribution of the last term [i.e., $\partial_\xi e^{\Delta c(j;\xi)}$] is of high order and consequently it can be omitted. The remaining part may be written according to the mean value theorem as:

$$\Delta c(i) = \rho_c \theta \int c(i,j;\xi_0) \left\{ -\beta_c v(i,j) e^{-\beta_c w(i,j)} + \beta_c \mu - 3/2 \right\} e^{\Delta c(j;\xi_0)} dj = \rho_c \theta \chi(i;\xi_0). \qquad (25)$$

Note that the value $\xi_0$ (contained between zero and one) depends on $\theta$.
Using equations (16) and (25) and the relation between the densities, $\rho_l$, $\rho_v$, and $\rho_c$ and the parameter $\theta$ (see for instance reference 6),

$$\rho_l - \rho_v \sim \rho_c \theta^\beta, \qquad (26)$$

where $\beta$ is the critical exponent which should not be confused with $\beta = 1/kT$ used in the beginning of this paper, the surface tension takes the form:

$$\gamma \sim (kT/4S)(\rho_l - \rho_v)^4 \theta^{4-2\beta}(z/z_c)^2 \iint \partial_{z_1} \chi(1;\xi_0) e^{\rho_c \theta \chi(1;\xi_0)} \partial_{z_2} \chi(2;\xi_0) e^{\rho_c \theta \chi(2;\xi_0)}$$

$$c(1,2)(r^2_{12} - z^2_{12}) \, d1 \, d2,$$

$$\sim \left\{ (kT/4) \theta^{4-2\beta}(z/z_c)^2 \iint \partial_{z_1} \chi(1;\xi_0) e^{\rho_c \theta \chi(1;\xi_0)} \partial_{z_2} \chi(2;\xi_0) e^{\rho_c \theta \chi(2;\xi_0)} c(1,2) \right.$$

$$\left. (r^2_{12} - z^2_{12}) \, dz_1 d\Omega_1 \, d\mathbf{r}_2 d\Omega_2 \right\} (\rho_l - \rho_v)^4, \tag{27}$$

which, at the first order, may be written in the same form as in equation (14). However the constant $K$ which remains a smooth function of the temperature is now, as it can easily be seen, more sensitive to this parameter.

This result is obtained whatever the value of the critical exponent $\beta$ is. It follows that the surface tension is given by the Macleod formula as a first approximation. The constant $K$ becomes very sensitive to the temperature near the critical point.

Clearly, the constant, $K$, is a function of the parameters $\theta$ and $\xi_0$ and therefore, it depends on the temperature of the system: *(i)* explicitly, because of the first parameter and, *(ii)* implicitly because of the second.

### 3. Concluding remarks

The purpose of this work was to justify Macleod's formula using the statistical-mechanical definition of the surface tension. This property was expanded in power series of the difference in densities, $(\rho_l - \rho_v)$. It takes the form of Macleod's formula when only the lowest order in its expansion is retained. The constant $K$ is however a weak function of the system temperature.

It is to be mentioned that:

*(i)* this result is general and so, it may be applied whatever the nature of the fluid is, if the condition of additive pairwise potential is satisfied.

*(ii)* Although the form of the first term of the surface tension expansion remains the same near the critical temperature, the constant $K$ becomes very sensitive to this parameter when it increases and tends to its critical value, $T_c$.

**Acknowledgment**

This research is supported by the Chemical Sciences Division Office of Basic Energy of the U.S. Department of Energy, GrantDE-FG02-84ER13229.